\documentclass[twocolumn,aps,floatfix,superscriptaddress,nofootinbib]{revtex4}
\pdfoutput=1 %This is for the ArXiv submission only
\usepackage{amsmath,amssymb,eucal,graphicx,float,epstopdf}
\usepackage{subfigure}
\usepackage{url}
\usepackage{float}
\usepackage{color}

\newcommand\gtwid{\mathrel{\raise.3ex\hbox{$>$\kern-.75em\lower1ex\hbox{$\sim$}}}}
\newcommand\ltwid{\mathrel{\raise.3ex\hbox{$<$\kern-.75em\lower1ex\hbox{$\sim$}}}}

\begin{document} 

\title{Optimal Storage for Solar Energy Self-Sufficiency}
\author{Anders E. Carlsson}
\affiliation{Department of Physics, Washington University, St. Louis, Missouri}
\author{S. Redner}
\affiliation{Santa Fe Institute, 1399 Hyde Park Road, Santa Fe, New Mexico, 87501}

\begin{abstract}
We determine the energy storage needed to achieve self sufficiency to a given reliability as a function of excess capacity in a combined solar-energy generation and storage system. Based on 40 years of solar-energy data for the St.\ Louis region, we formulate a statistical model that we use to generate synthetic insolation data over millions of years. We use these data to monitor the energy depletion in the storage system near the winter solstice. From this information, we develop explicit formulas for the required storage and the nature of cost-optimized system configurations as a function of reliability and the excess of generation capacity. Minimizing the cost of the combined generation and storage system gives the optimal mix of these two constituents.  For an annual failure rate of less than 3\%, it is sufficient to have a solar generation capacity that slightly exceeds the daily electrical load at the winter solstice, together with a few days of storage.
\end{abstract}

\maketitle

Moving away from fossil fuels to renewable energy is a crucial step to minimize the extent of global warming. Because renewable energy sources, such as wind and solar, are intermittent, achieving a 100\% renewable scenario requires either a large excess generation capacity, a substantial amount of storage, or a judicious mixture of the two. Understanding the nature of is tradeoff between excess capacity and storage is crucial for the design and optimization of effective renewable energy systems.  Understanding the factors that determine the tradeoff will improve our grasp of the right balance between the uncertain costs of generation and storage in the future.

This tradeoff is characterized by two fundamental parameters: the \emph{generation factor g}, the ratio of the average annual generation capacity to the annual load, and the storage capacity $S$, the number of days of electrical load that reside in a storage system. Various simulation studies have given scattered values for the optimal mix of $g$ and $S$ values, with little understanding of how they depend on physical parameters. Ref.~\cite{CaldeiraEES18} examined combined energy generation and storage systems in the US and in specific subregions, with both wind and solar generation. For solar-only generation, a system with $g \approx 2.1$ and a storage equivalent $S$ of 4 days of load was virtually 100\% reliable, defined as the fraction of total energy demand that was met by renewables plus storage.  However, equally high reliability was obtained with $g\approx 1.3$ and a month of storage. Ref.~\cite{BofingerRE10} focused on the case $g=1$, with a mix of solar and wind energy and found that a storage $S$ of 1.1--2.5 months was required.

Ref.~\cite{tong2020effects} developed optimized energy systems for the continental US over a range of storage costs, using the same underlying model as in Ref.~\cite{CaldeiraEES18}. For inexpensive storage, they found  $g\simeq 2.2$, while more expensive storage required an increase in the capacity to $g\simeq 2.7$. Concomitantly, the amount of storage dropped from about 5 days of load to 1 day. Ref.~\cite{KemptonJPS13} determined optimal solutions for a power network in the eastern US with disparate storage modalities. The requisite $g$ values were in the range of 2.5--2.9 and $S$ between 0.3--3 days, depending on the type of storage.  Related studies~\cite{HoffmannRE11,GreinerEP12,JacobsonPNAS15,PohlJES17} added easily dispatchable renewable sources, such as hydroelectric power, which reduced the required storage.

Given the range of the these predictions about optimal configurations, a need exists for an analytical theory that would: (a) clarify the relation between input physical parameters and the performance of a combined generation/storage system, and (b) help constrain the parameters of this system to guide the realm of feasibility. In this work, we construct such a theory that is based on an idealized, but general model that faithfully incorporates the actual solar irradiation statistics, including seasonality and day-to-day correlations. This theory allows us to specify the nature of an optimal generation/storage system and make explicit predictions about its cost and reliability. Although optimal systems will in general include both wind and solar energy, we treat only solar energy in order to obtain a theoretically tractable model. We believe that the general features of our results will hold for mixed systems as well.

Our model extends previous analytical theories based on simplified solar irradiation statistics. Ref.~\cite{ZoglinSC86} assumed a deterministic day-night profile, while ignoring daily and seasonal fluctuations. Refs.~\cite{BucciarelliSE84,BucciarelliSE86,GordonSE87} included daily, but not seasonal weather variations, and day-to-day correlations in some cases. They found that the failure probability decays exponentially with increasing storage capacity, and Ref.~\cite{GordonSE87} gave explicit formulas for the storage capacity required to achieve a given reliability. Ref.~\cite{markvart1996sizing} included the effects of seasonality in generation and/or load but did not treat random weather fluctuations. Ref.~\cite{LorenzoSEM92} used an empirical fit to reliability simulations based on historical weather data (including seasonality), and found an exponential relationship between generation and storage. However, a principled theory that quantitatively treats the combination of stochastic daily weather fluctuations, day-to-day correlations, and seasonality does not yet seem to exist.

We begin by first outlining basic features of the solar-flux data for the St.\ Louis region, which typifies those of the entire US. We then introduce our data-driven model and use it to develop analytic formulas for the failure rate and storage capacity needed to achieve a given reliability. We use these to calculate the generation and storage capacities of a combined system that minimizes the cost and yet is extremely reliable. We verify our predictions based on simulations of millions of years of synthetic data. 

\section{Empirical Background}
\label{sec:empirical}
\subsection{Solar Flux Data}
\label{subsec:data}

\begin{figure}[ht]
\centerline{\includegraphics[width=0.35\textwidth]{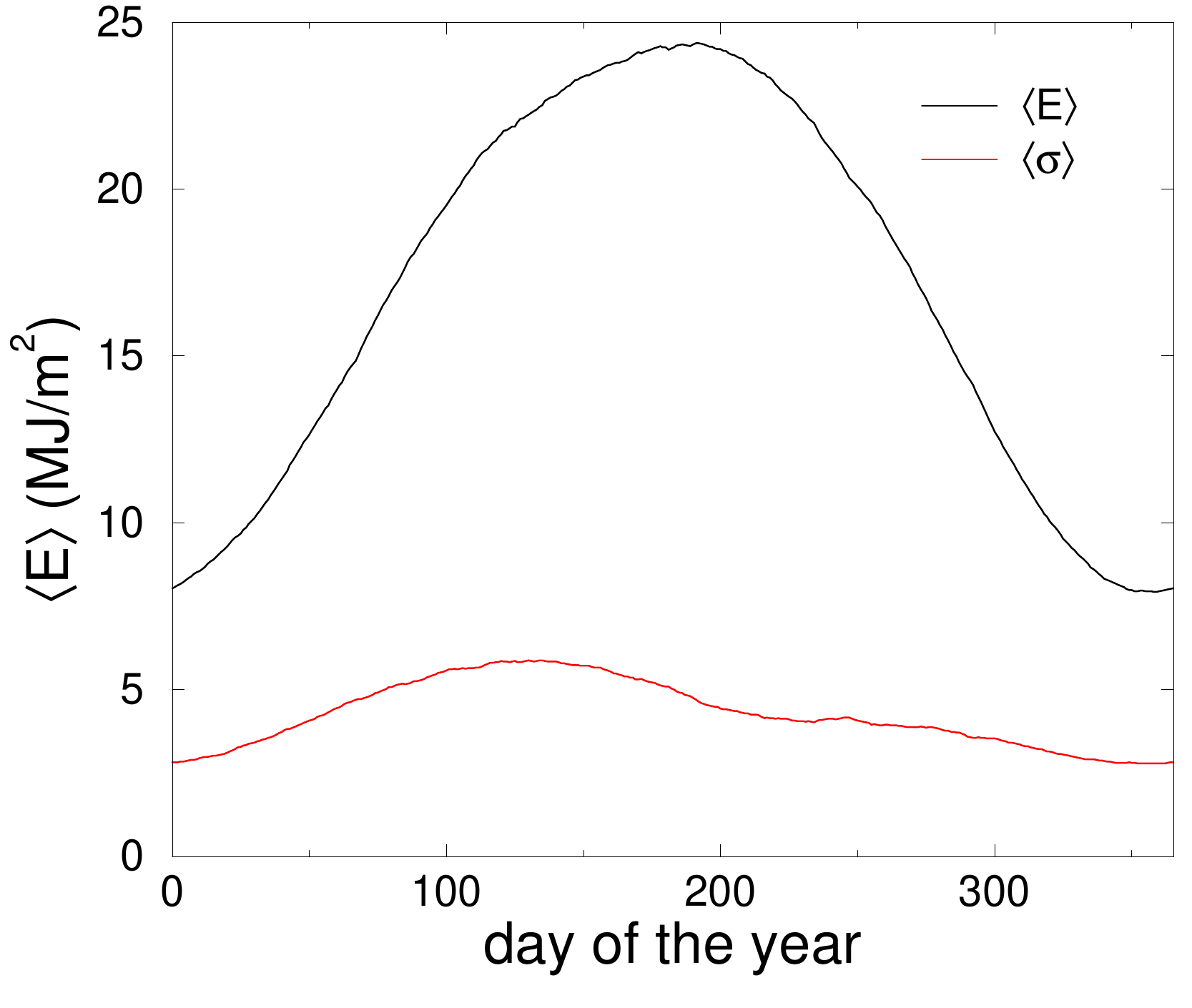}}
\caption{Average daily energy $\langle E\rangle$ per unit area from 1980--2019 on the St.\ Louis region (black), and the daily standard deviation $\langle\sigma\rangle$ in this quantity over the same period (red).  These data are smoothed by averaging over 45 consecutive days. }
\label{fig:av-E}  
\end{figure}

\begin{figure}[ht]
\centerline{\includegraphics[width=0.375\textwidth]{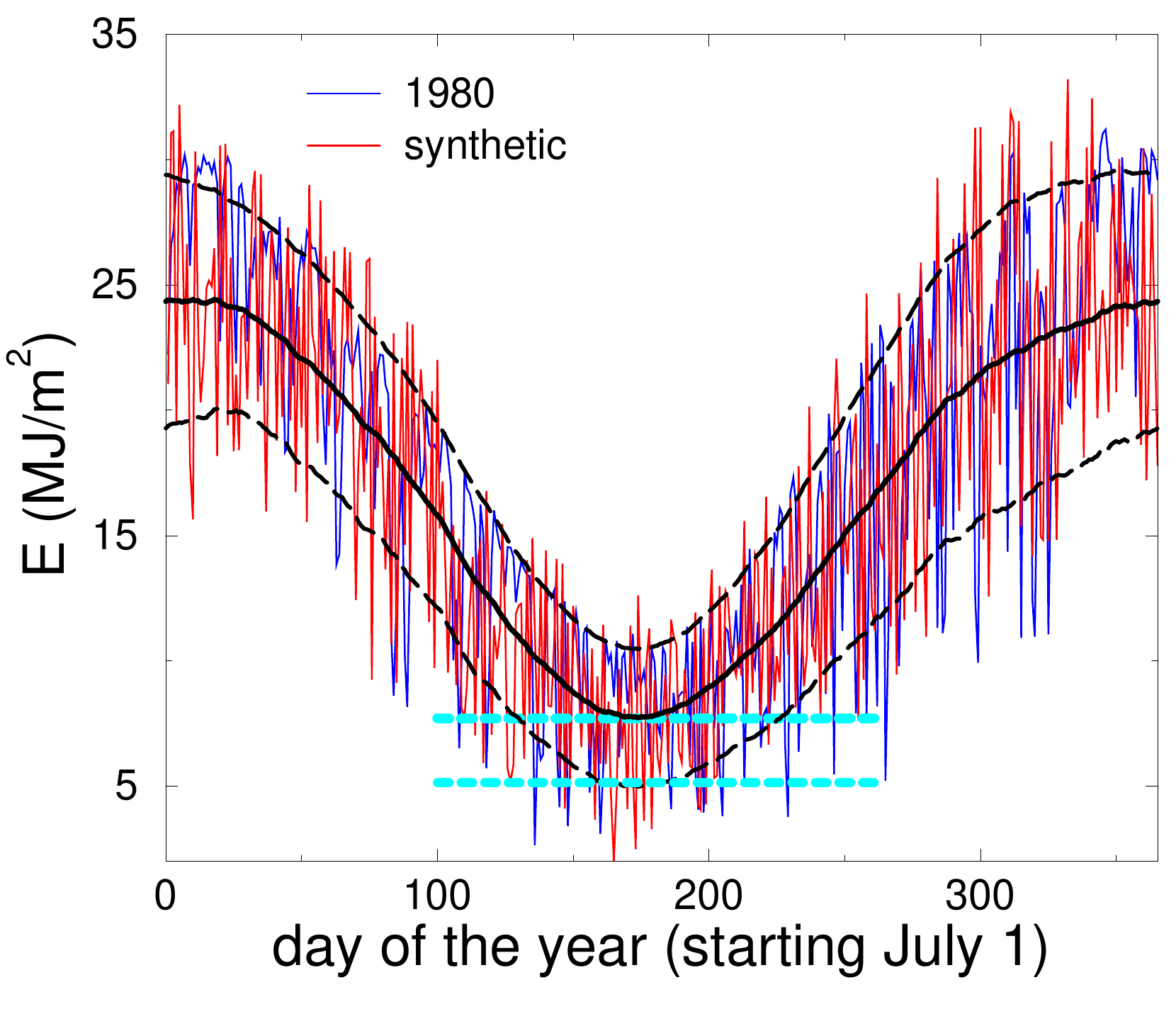}}
\caption{Insolation fluctuations. The black curve is smoothed 45-day average daily energy per unit area on the St.\ Louis region, 1980--2019  and the one standard deviation range is dashed curve. Also shown is the daily data for 1980 (blue) and a typical realization of our synthetic data (red).  The upper dashed blue line corresponds to the load with $f\!=\!1$ (see text for the definition of $f$) and the lower corresponds to $f\!=\!1.5$. The difference between the daily data and the dashed lines gives the daily energy surplus/deficit for these two $f$ values.
}
\label{fig:1}  
\end{figure}

To illustrate the issues and as a preliminary to develop our model, we first present and analyze data for the solar flux on a 270 km $\times$ 270 km region centered on St.\ Louis over the 40-year period 1980--2019. This region is large enough that its energy needs can be met by covering a small fraction of the total land area with solar panels, but small enough that power transmission across the region is nearly lossless and instantaneous.  The solar data, from the MERRA-2 dataset~\cite{BacmeisterGMD15}, is in the form of the energy flux for each hour of the day from 1980--2019 (see Appendix \ref{sec:SM-ws}). We determine the daily incident energy per unit area by multiplying each hourly energy flux by the number of seconds in an hour, and then adding these values over a single day.  This gives an average daily solar energy per unit area that ranges between roughly 8--25 MJ/m$^2$ from the winter minimum to the summer maximum, with daily extrema of 1.53 MJ/m$^2$ and 32.1 MJ/m$^2$ over the 40 years of data (Figs.~\ref{fig:av-E} and \ref{fig:1}). For simplicity in our analysis, the data for February 29 in leap years are dropped, so that our results are based on the 40-year period 1980--2019 in which all years consist of 365 days.

The average daily solar energy is roughly sinusoidal, with the maximum at day 200 (July 18, nearly one month after the summer solstice) and the minimum at day 357 (December 23, a few days after the winter solstice). The standard deviation in the daily solar energy also has a systematic time dependence that ranges between 1.5--5.5 MJ/m$^2$, with maximal fluctuations occurring in the early spring. Near the winter solstice, the magnitude of the fluctuations is about 35\% of the mean value. On the minimum-insolation day, $ E \approx 7.95$ MJ/m$^2$ and $\sigma\approx 2.79$ MJ/m$^2$.  These numbers, which will play a central role in our ensuing analysis, are based on averaging the daily energy data over a 45-day window. 

To illlustrate the influence of fluctuations on the daily insolation data, Fig.~\ref{fig:1} shows the daily solar flux on the St.\ Louis region for the single year 1980.  For later convenience in our theoretical and numerical modeling, the time origin has been shifted so that the year begins on July 1.  The basic feature of these data is that daily solar fluctuations significantly perturb the average annual cycle.  Around the winter solstice, which is the most crucial time of the year for the reliability of a combined solar generation and storage system, the minimum solar insolation is roughly a factor 5 less than the \emph{average} maximum solar insolation. If the storage system is nearly depleted near the winter solstice, multiple overcast days can quickly lead to a system failure.  Thus day-to-day fluctuations in solar flux play an important role in determining the optimal tradeoff between generation and storage.

\subsection{Generation and storage costs}
\label{generationandstorage}

Our determination of the optimal system configuration is based on two key costs: the cost $C_g$ of the generation capacity to supply the daily electrical energy load $L$ at the winter solstice (based on the average insolation on that day), and the cost $C_s$ of energy storage to cover one day of electrical load. This daily load of the St.\ Louis region is\footnote{This is obtained from the continental US yearly electricity consumption of 4$\times 10^{12}$ kWh   (https://www.statista.com/statistics/201794/us-electricity-consumption-since-1975/), dividing by 365 to get consumption per day and then by 100 (taking the St.\ Louis region to contain about 1\% of the US population)} roughly $4 \times 10^{14}$\,J, or $1.1\times 10^8$\, KWh. This corresponds to an average power usage of 4.6 $\times 10^6$\, kW. 

It is conventional to express the cost of solar panels in dollar per watt.  Using the current solar panel cost of \$1.50/W~\cite{cost-solar}, the cost of generation is thus $C_g\approx$ \$75 billion. This cost grows roughly linearly with the area of the solar farm\footnote{There is little economy of scale for a large solar farm~\cite{scale}. While the cost per solar panel decreases as the number of installed panels increases, there are additional costs associated with transmitting solar power from the farm to end users. These transmission costs largely negate the installation economy of scale; such costs do not exist for rooftop solar panels for household use.}. For 20\% efficient solar cells (close to the best that are currently available~\cite{spe}), the required solar farm area is $4 \times 10^{14} \text{J}/(0.20 \times 7.95\times 10^6 \text{J}/\text{m}^2) \approx 2.5 \times 10^8$\,m$^2\equiv A_0$. This roughly corresponds to a 16 km $\!\times\!$ 16 km square. The cost of a solar farm of area $fA_0$ will therefore be $f\,C_g$.  The excess generation capacity, $(f-1)L$, is a fundamental metric of the generation system. Using the current price of \$1.25/m$^2$~\cite{land} for rural land in the region, the land cost of the solar farm is approximately \$300 million; this is negligible compared to the solar panel costs and will be ignored. 

The cost to store one day of electrical energy load for the St.\ Louis region at the current price of \$200/KWh is $C_s = 1.1 \times 10^8 \text{kWh} \times \$200/\text{kWh} \approx$ \$22 billion~\cite{ziegler2019storage}.  It is convenient to measure the capacity $S$ of the storage system in units of the daily electrical energy load in the St. Louis region. We define a storage system of capacity $S$ as one that can supply $S$ days of electrical load to this region. The cost of this storage system therefore is $C_s S/L$.

Since roughly 60\% of a 24-hour period is dark at the winter solstice in the St.\ Louis region and total electrical energy use is roughly time independent in the winter~\cite{constant1,constant2}, there is a baseline storage need of 60\% of the daily load to cover the energy use when it is dark. If there were no day-to-day fluctuations in the solar flux, this baseline storage, together with the solar energy gathered during the day by a solar farm of area $A_0$ could fully supply the regional electrical energy needs during a 24-hour period at the solstice, and thus throughout the year.

The existence of insolation fluctuations has several essential consequences. First, the optimal area of the solar farm must be larger than $A_0$ and the storage capacity must be larger than the 60\% of daily energy use that is needed to deal with the regular diurnal fluctuations.  Second, we will see that it is impractical to achieve 100\% reliability with this combined solar generation and electrical storage system. Thus it is necessary to balance the tradeoff between reliability and cost. Establishing how generation capacity and storage combine to achieve a given reliability, and understanding the tradeoff between reliability and cost, are primary goals of this work. We will find that the optimal cost system configuration is determined by the ratio of storage to generation costs, $C_s/C_g$. The above numbers give roughly 0.3 for this ratio. Since storage costs are rapidly decreasing~\cite{ziegler2019storage}, we will explore the consequences of potential future storage cost reductions by up to a factor of 7.

\section{Synthetic Data and Simulations}

Because of the substantial day-to-day fluctuations in the solar flux, the 40 years of available data are too sparse to determine the reliability of a combined solar farm/storage system with statistical significance.  To formulate a generally applicable theory, we first construct synthetic daily insolation data that faithfully incorporates the annual trends, the daily fluctuations, and the day-to-day correlations that are present in the solar flux data for the St.\ Louis region.  The simple and direct algorithm that we use to construct these data allows us to readily generate time series for millions of years. From these, we obtain statistically meaningful results about the reliability and cost of a combined solar power generation and storage system.

\begin{figure}[ht]
\centerline{
\subfigure[]{\includegraphics[width=0.245\textwidth]{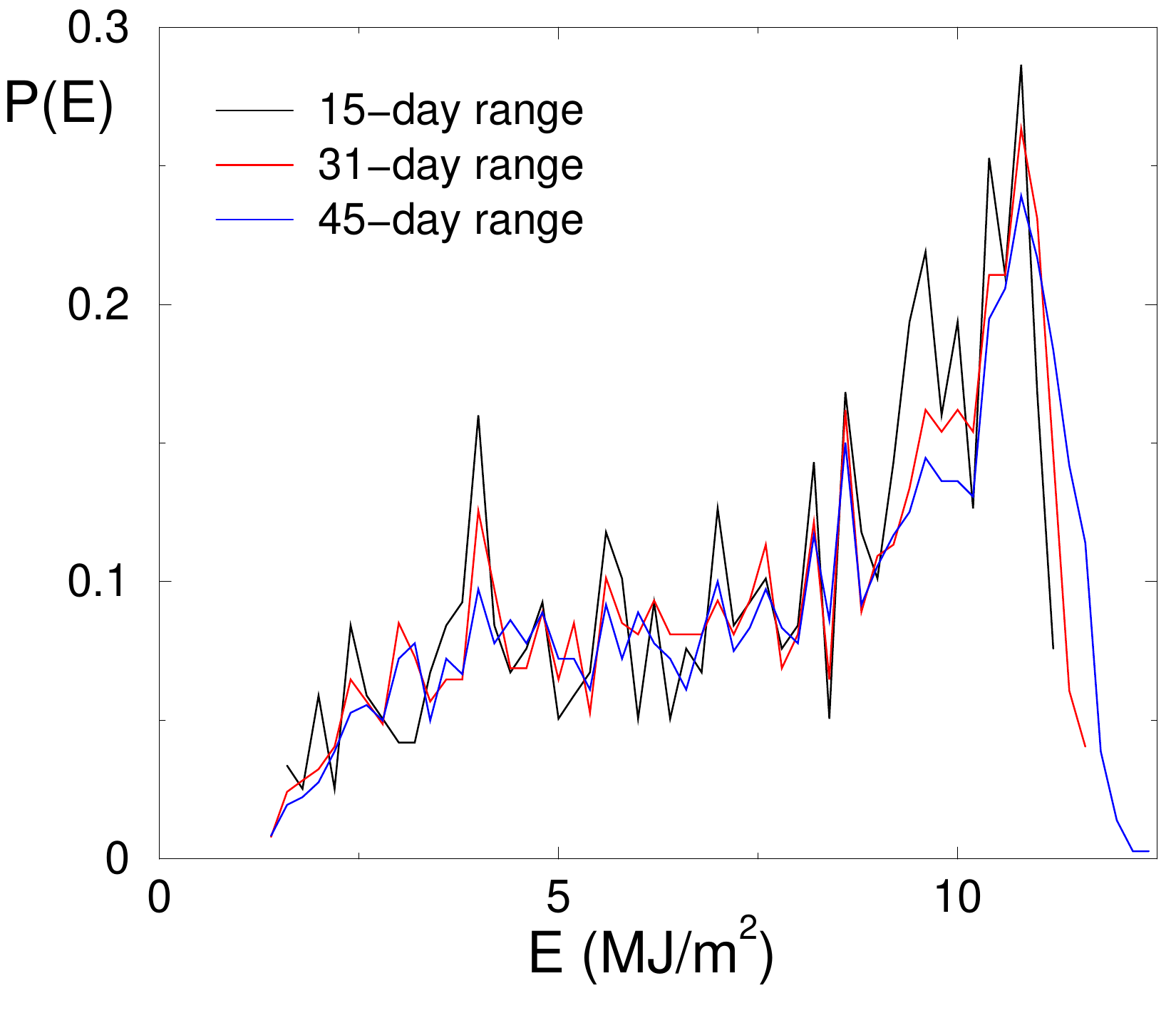}}
\subfigure[]{\includegraphics[width=0.245\textwidth]{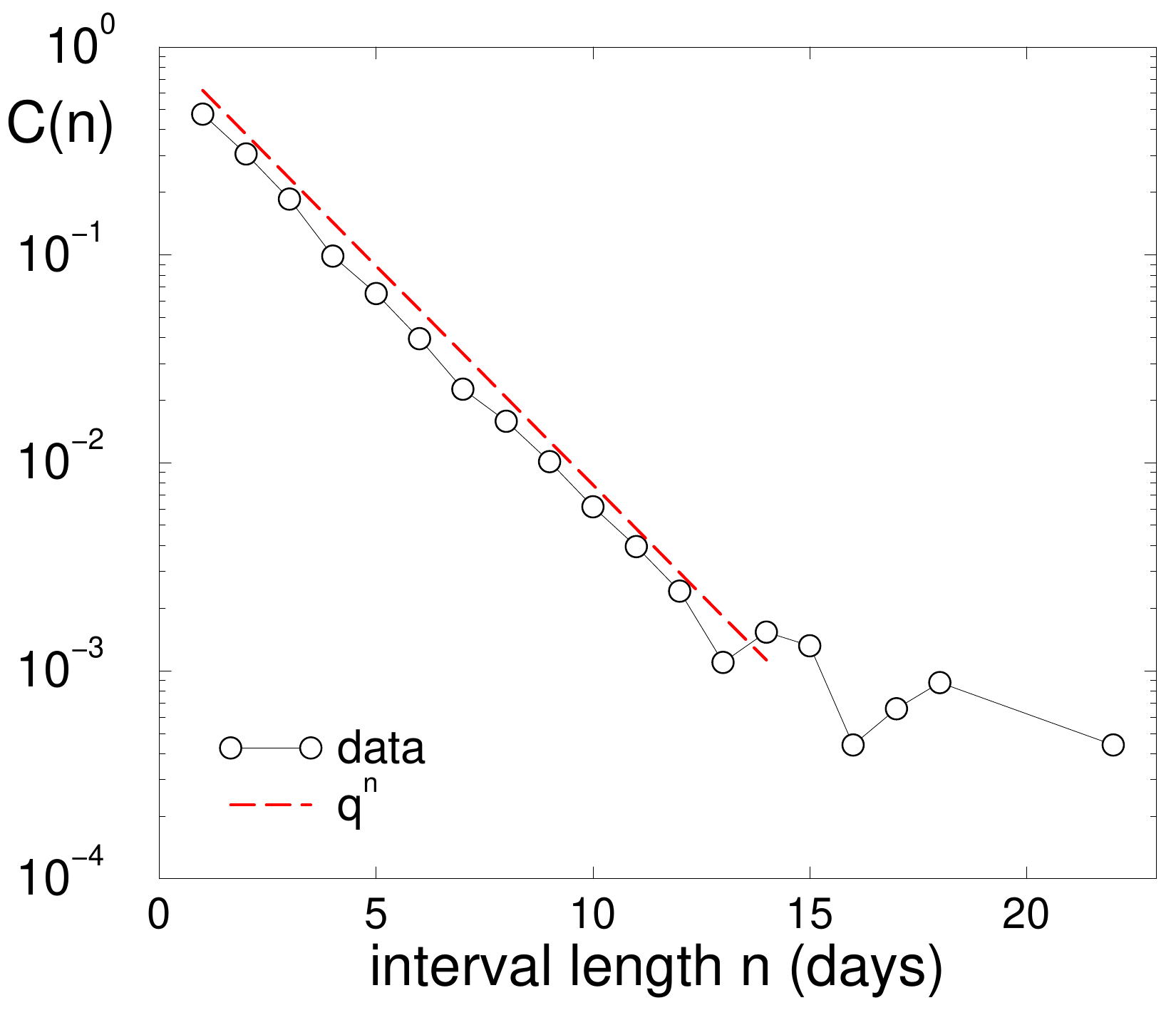}}}
\caption{(a) The probability distribution $P(E)$ for the daily energy per unit area over symmetric time intervals around the minimum-energy day (day 357). (b) Probability $C(n)$ that the ratios $r_j\equiv E_j/\langle E_j\rangle$ are all greater than or all less than 1 over $n$ consecutive days.  Also shown is the exponential best fit to the  data in the range $n\leq 13$, $C(n)\propto q^n$, with $q=0.6157$.}
\label{fig:seasonal-E}
\end{figure}

To construct the synthetic data, we require two additional features beyond the average daily incident energy and its standard deviation: (a) the distribution of energy for each day of the year, and (b) the day-to-day energy correlations. The energy distributions away from the winter solstice are irrelevant when $f > 1$ because there will be ample solar energy plus stored energy to meet the daily load on any given day that is not near the solstice.  It is only near the winter solstice that the daily energy distributions become relevant.  However, 40 years of data are too sparse to accurately represent these distributions. To obtain daily energy distribution data of reasonable quality, we aggregate these distributions over symmetric time ranges of 15, 31, and 45 days around the minimum solar-energy day (day 357). These distributions are nearly the same for the three time ranges (Fig.~\ref{fig:seasonal-E}(a)); this justifies using a universal shape for the daily energy distribution near the winter solstice. For simplicity, we replace the actual and somewhat triangular-shaped  distribution by a uniform distribution whose width is chosen to be the same as that of the data.

There are also day-to-day correlations in the energy flux that reflect the well-known feature that the weather on consecutive days is more likely to be similar than different~\cite{BucciarelliSE86}.  To quantify these correlations, we start with the 40-year sequence of normalized daily energies $\{r_j\equiv E_j/\langle E_j\rangle\}$, where $E_j$ is the energy per unit area on the $j^{\text th}$ day of the year and $\langle E_j\rangle$ is its average, with $j$ ranging\footnote{There is a small error in the correlation function because we drop the data for the 10 leap days.} from 1 to 14600 ($40\times 365$).  We first determine the length of strings of consecutive days for which the ratios $r_j$ are either all greater than 1 or all less than 1.  We then obtain the probability distribution $C(n)$ for the number of consecutive days $n$ where all the $r_j$ are greater than 1 or less than 1.

In the absence of correlations in the daily solar flux, the string length distribution would decay in $n$ as $C(n)=(1/2)^n$.  However, the actual correlations decay as $q^n$, with $q\approx 0.6157$ over the range of 1--16 days (Fig.~\ref{fig:seasonal-E}(b)).  Beyond 16 days, the correlations decay more slowly still.  However, the frequency of such long strings of 16 days or longer is roughly once every 6 years.  In generating our synthetic data, we ignore these extremely rare events and use the simple exponential decay $C(n)\propto q^n$ for all $n$.

It is convenient to shift the time origin so that the year begins on July 1. We first determine the solar energy on July 1 (now day 1).  This energy is given by a random number that is uniformly distributed in the range $[E_1\!-\!\sqrt{3}\,\sigma_1,E_1\!+\!\sqrt{3}\,\sigma_1]$, where $E_1$ is the average energy and $\sigma_1$ is the standard deviation on July 1 (see Fig.~\ref{fig:av-E}).  The factor $\sqrt{3}$ ensures that the standard deviation that arises from the uniform distribution matches that of the actual data. For each successive day $j$, the solar energy is again chosen uniformly from the range $[E_j\!-\!\sqrt{3}\,\sigma_j,E_j\!+\!\sqrt{3}\,\sigma_j]$, but augmented by the following construction to incorporate the energy correlations. The sense of the deviation from the average energy on the current day $j$ (either $E_j/\langle E_j\rangle>1$ or $E_j/\langle E_j\rangle<1$) is the same as the sense of the deviation on the previous day with probability $q$.  This persistent random-walk construction~\cite{weiss1983random,weiss1994aspects} ensures that the string length distribution asymptotically decays as $q^n$, as in Fig.~\ref{fig:seasonal-E}(b). The synthetic data accurately mimic both the annual variation, as well as the day-to-day fluctuations of the incident energy, as illustrated by a typical realization of synthetic daily energies in Fig.~\ref{fig:1}. 

\begin{figure}[ht]
\centerline{\includegraphics[width=0.35\textwidth]{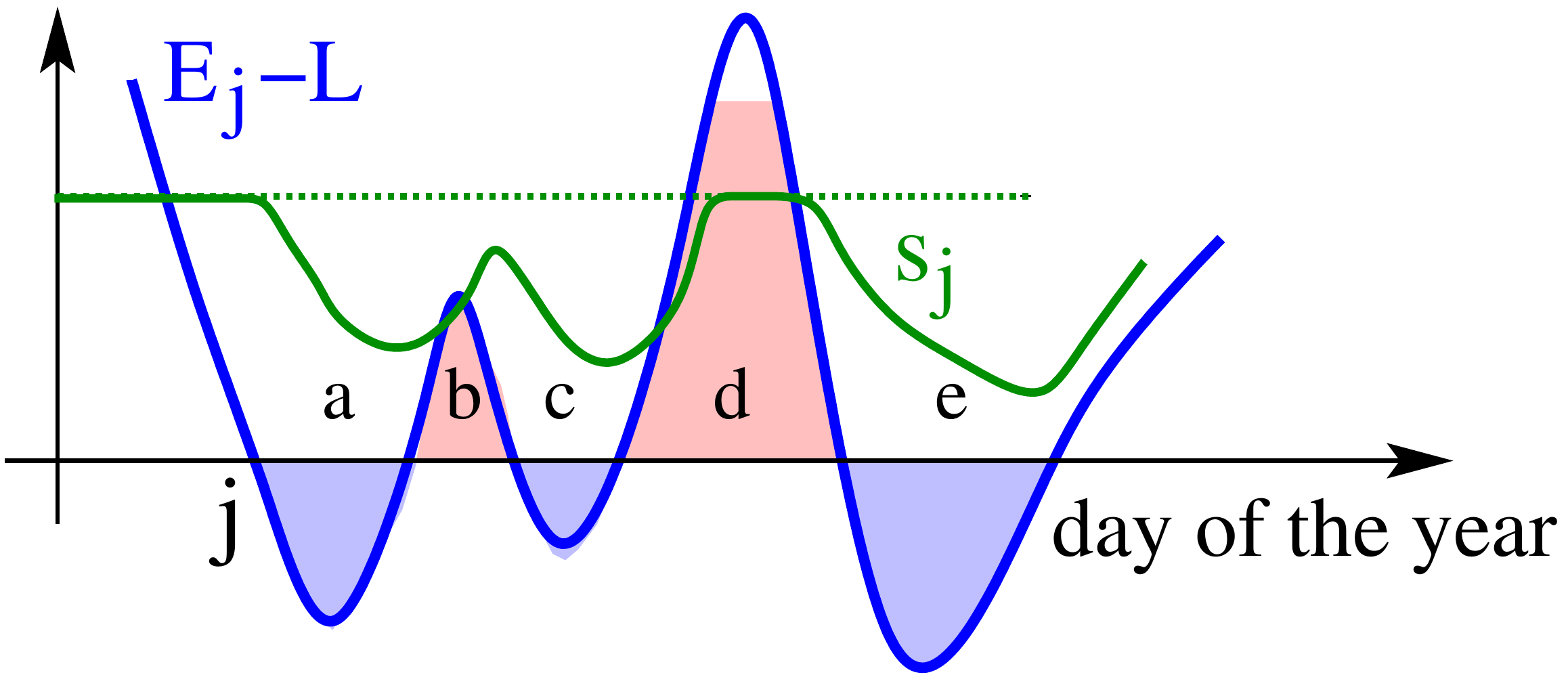}}
\caption{Schematic and not to scale dependence of the daily energy minus the load near the winter solstice (blue curve), with three periods of below average insolation ($a,c,e$) and two above-average periods ($b,d$). The extent of the energy deficits and surpluses are shown by the blue and red shaded areas.  The green curve indicates the instantaneous storage $s(t)$ and the green dotted line indicates full storage.}
\label{fig:fluct}  
\end{figure}

With this computational approach, we generate millions of years of synthetic insolation data over a two-dimensional mesh of thousands of $(f,S)$ values. We start with a full storage system, that is, $s=S$ on July 1. For a given pair $(f,S)$, the stored energy $s_j$ on the $j^\text{th}$ day of the year is a random variable that changes daily according to $s_{j+1} = s_j + E_j-L$, where $L$ is the daily load, subject to the constraint that $s_j$ can never exceed $S$ (Fig.~\ref{fig:fluct}). Since the model treats only the total energy in a day, it does not include the diurnal variation mentioned in Sec.~\ref{generationandstorage} that will require an additional constant storage requirement of $0.6L$.  

The time evolution in the model defines a biased random-walk-like process on the interval $[0,S]$, in which the bias corresponds to the difference between the daily insolation and the daily load, and the day-to-day insolation fluctuations correspond to random noise. System failure occurs whenever $s_j$ reaches zero. The failure probability $\varepsilon$ is defined as the fraction of years for which failure occurs. The cost-optimized system configuration is defined as the pair of $(f,S)$ values that has the lowest cost and a failure probability less than a specified value.

\section{Generation/Storage Tradeoff}
\label{sec:synth}

Due to fluctuations in daily insolation, even a system with $f > 1$ will be insufficient to supply the electrical load unless storage is included. We will construct a theory to determine the range of possible mixes of generation and storage that achieve a given reliability. We treat only the constraints that arise from storage-capacity limitations and not from power-delivery limitations.  We also assume 100\% efficient storage, perfect power transmission across the region, and a constant daily load.  

The stylized time history of the insolation and stored energy near the winter solstice (Fig.~\ref{fig:fluct}) also illustrates the tradeoffs involved in optimizing the combined system. In this figure, the energy deficit during period $a$ is larger than the surplus in the following period $b$. Thus full storage on the $j^{\rm th}$ day is only partially replenished in period $b$.  Conversely, while the storage is fully replenished in period $d$ with a large solar surplus, some of this surplus is wasted because of the limited storage capacity (indicated by the cutoff in the red area). The optimal storage system should maximize the energy returned to storage during surplus days near the winter solstice, while minimizing cost.

\subsection{Stored energy distribution}
\label{subsec:stored}

For a solar farm of area $A_0$, insolation tends to replenish the storage during most of the year; this time range corresponds to what we term the \emph{strong-bias} regime. Conversely, for an average day near the winter solstice, the insolation roughly matches the load, so that the state of the storage system change only slightly from day to day. We term this time range as the \emph{weak-bias} regime. In an optimal design, the storage system should be nearly depleted through the winter solstice.  Otherwise, excess unused storage capacity exists that increases the system cost without meaningfully increasing its reliability.

For each day of the year, there is a day-specific average distribution of energy in the storage system.  We will determine these daily distributions over a period around the winter solstice.  From these distributions, we will determine how the annual failure probability $\varepsilon$ depends on $f$ and $S$. Sec.~\ref{sec:cost} uses these relations to calculate the necessary storage and generation capacity in a cost-optimized system configuration. 

To compute the stored energy distribution on a single day, we first treat the idealized situation of a strong and time-independent bias. Based on the biased random-walk picture described above, the distribution of stored energy on the $j^{\text{th}}$ day of the year, $P_j(s)$, attains the steady-state form 
\begin{align}
\label{Ps}
   P_j(s)= \mathcal{N}\big(e^{\lambda_j s}-1\big)\,,
\end{align}  
with normalization constant $\mathcal{N}=\lambda_j/(e^{\lambda_j S}-1-\lambda_j S)\approx \lambda_j \, e^{-\lambda_j S}$ determined by requiring that $P(s)$ integrates to 1. A salient feature of Eq.~\eqref{Ps} is that the decay constant in the exponential is different for each day of the year. (Related approaches for this distribution were given in Refs.~\cite{BucciarelliSE84,BucciarelliSE86,GordonSE87}.) 

To begin, we determine the decay constant $\lambda$ for the case where the bias is fixed. Then we incorporate the effect of a seasonal variation in this bias, as well as the role of day-to-day correlations in the insolation, to find the decay constants in the storage distribution for a range of days about the winter solstice.  From these results, we will compute the annual failure probability. 

When the bias is constant, the stored energy after each day changes by the average solar energy surplus (or deficit), $(f-1)L$, plus or minus a uniform random variable in the range $[-\sqrt{3}f\,\widetilde{\sigma}\, L,\sqrt{3}f\,\widetilde{\sigma}\, L]$ to account for day-to-day fluctuations. Here $\widetilde{\sigma}=\sigma/E$ is the ratio of the standard deviation in the energy density to the energy density on any given day (see Fig.~\ref{fig:av-E}). Thus the governing equation for $P(s)$ is 
\begin{align}
P(s) = \frac{1}{2\sqrt{3}f\,\widetilde{\sigma}\, L} \int_{s_{-}}^{s_{+}}P(s')\,ds'\,,
\label{dynamic}
\end{align}
where $s_{\pm}=s-(f-1)L\pm \sqrt{3}f\widetilde{\sigma} L$.  This equation is satisfied by the exponential form of Eq.~\eqref{Ps}, where the decay constant satisfies
\begin{align}
\frac{\sinh{(\lambda_{\rm cb} {\sqrt{3}f\widetilde{\sigma} L})}}{\lambda_{\rm cb} {\sqrt{3}f\widetilde{\sigma} L}} = \exp{[\lambda_{\rm cb}(f-1)L]}.
\label{sinh}
\end{align}
Here we write the decay constant as $\lambda_{\rm cb}=\lambda_{\rm cb}(f)$, with subscript cb to emphasize that we specialize to the constant-bias case. Equation~(\ref{sinh}) gives the dimensionaless quantity $\lambda_{\rm cb}L\to 2(f-1)/(f^2\widetilde{\sigma}^2)$ for $f\to 1$, while $\lambda_{\rm cb}$ deviates slightly from linearity for larger $f$ (Fig.~\ref{fig:sinh} in Appendix~\ref{sec:SM-linear}). Over the practical range of $1<f<1.5$, this dependence is accurately described by a linear interpolation between $\lambda_{\rm cb}=0$ at $f=1$ and $\lambda_{\rm cb}L\approx 5.675$ that arises by numerically solving Eq.~\eqref{sinh} at $f=1.5$. Thus we infer $\lambda_{\rm cb}L= \Gamma\,(f-1)$, with $\Gamma \approx 11.35$. As we shall see, this linear interpolation allows us to construct an analytical theory for the failure probability that incorporates both seasonality and fluctuation effects.

Seasonality causes the steady-state distribution of stored energy to be slightly different for each successive day of the year; thus we now write this distribution as $P_j(s)$, with $j$ indexing the individual day. We first determine $P_{\rm min}(s)$ on the minimum-insolation day, with $\lambda_{\rm min}$ the decay rate on this day. This decay rate would equal 0 when $f \to 1$ within the above constant-bias description. However, our simulations show that the storage distribution still has a nearly exponential form even when $f=1$ (see Appendix~\ref{sec:SM-storage}). Thus we need to postulate a functional form for the decay constant on the minimum-insolation day that interpolates smoothly between the limiting cases of a value $\lambda_0$, which we will determine, when $f=1$ and $\Gamma(f-1)$ when $f-1$ is not small. A simple form that satisfies these criteria is
\begin{align}
\lambda_{\rm min} (f) = \tfrac{1}{2}\big[\Gamma(f-1)+\sqrt{(4\lambda_0^2+\Gamma^2(f-1)^2}\big]\,.
\label{lambdamin}
\end{align}

We also need the steady-state storage distributions $P_j(s)$ and their associated decay rates $\lambda_j(f)$ on a range of days around the minimum-insolation day. To obtain these distributions, we use the fact that the average daily generated solar energy $E_j$ on days near the winter solstice is well described by the quadratic $E_j/L= f+f(j-j_{\rm min})^2/\tau^2$, where $j_{\rm min}$ is the day of minimum insolation and $\tau= 72$ (in units of days) is determined by fitting to the 40-year average insolation data. For each day, the effect of the additional bias as one moves away from the minimum-insolation day is equivalent to increasing $f$ by $f(j-j_{\rm min}^2)/\tau^2$. Thus the day-specific decay constant is 
\begin{equation}
\label{lambdatf}
\lambda_j(f)=\lambda_{\rm min}(f) +f\,\Gamma\;(j-j_{\rm min})^2/\tau^2\,.
\end{equation}

Finally, we need to account for correlations in the daily insolation. To include these effects, we perform stochastic simulations of a system with constant bias $f=1.5$ (taken to be typical of the high-$f$ regime) and constant $\widetilde{\sigma} = 0.351L$. Logarithmic plots of $P(s)$ obtained both with and without correlations confirm the exponential behavior of $P(s)$, and show that including correlations reduces the decay parameter by $11$\%. Thus we take $\Gamma = (0.89) \times (11.35/L)=10.1/L$.

\subsection{The failure probability}

From the distribution of storage for each day of the year, we now determine the annual failure probability $\varepsilon$ of the storage system. We first estimate the failure probability $\varepsilon_j$ on each day $j$, and then add these daily failure probabilities over a time range that includes the winter solstice, to obtain the annual failure probability.  

The day-specific failure probability for the $j^{\text th}$ day of the year in the strong-bias limit is
\begin{align}
\label{failure}
   \varepsilon_j = \frac{1}{2\sqrt{3}f\widetilde{\sigma}_j L} \int_0^{\delta s} P_j(s) (\delta s -s)ds = A_j\,e^{-\lambda_j(f) S}\,.
\end{align}
That is, we integrate the storage distribution $P_j(s)$ over the energy range $\delta s = \sqrt{3}f\widetilde{\sigma}_j L-(f-1)L$, for which for the storage system can be completely depleted within a single low-insolation day, multiplied by the probability $(\delta s-s)/2\sqrt{3}f\widetilde{\sigma}_j L$ that starting with stored energy $s$, a negative energy step actually depletes the storage system.  The expression for $A_j$ is written in the SM and $\lambda_j(f)$ is the day-specific decay constant in Eq.~\eqref{lambdatf}.

To calculate the annual failure probability $\varepsilon$, we the sum the daily failure probabilities in Eq.~\eqref{failure} over the range of days where the quadratic dependence of the decay rate in Eq.~\eqref{lambdatf} applies, under the assumption that these daily failure probabilities are all independent. Because of the quadratic time dependence of $\lambda_j(f)$ in \eqref{lambdatf}, we convert the sum over a finite range of days $j$ around the insolation minimum to the following Gaussian integral over an infinite time range (see Appendix~\ref{sec:SM-fp} for details), in which days far from the minimum give negligible contributions:
\begin{align}
    \varepsilon =\sum_j \varepsilon_j
    &\simeq\int_{-\infty}^{\infty}  A(t)\, e^{-\lambda(t,f)S}\,dt \nonumber \\
    &=\frac{B}{\sqrt{\lambda_{\rm min}(f) S}}\;e^{-\lambda_{\rm min}(f)\,S},
    \label{epsfinal}
\end{align}
where we replace the index $j$ by the continuous time $t$, and $B$ is defined in Appendix \ref{sec:SM-fp}. 

We now invert this expression to solve for the required storage as a function of the reliability $\varepsilon$. In Appendix~\ref{sec:SM-S}, we show that the following approximate expression accurately describes the dependence of $S$ on $f$ and $\varepsilon$: 
\begin{align}
\label{S-opt}
S(f,\varepsilon) &=\frac{ \ln \left(\varepsilon_0/\varepsilon  \right)}{\lambda_{\rm min}(f)}\nonumber \\
&= \frac{2 \ln \left(\varepsilon_0/\varepsilon  \right)}{[\Gamma(f-1)+\sqrt{(4\lambda_0^2+\Gamma^2(f-1)^2}]} \,,
\end{align}
where $\varepsilon_0$ is defined in Appendix~\ref{sec:SM-S}.

\begin{figure}[ht]
  \centerline{\includegraphics[width=0.4\textwidth]{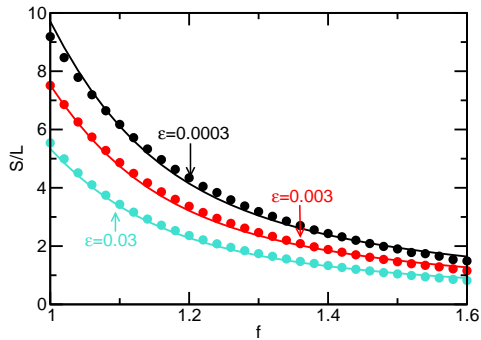}}
  \caption{Dependence of the storage $S$ needed for a
    failure probability  $\varepsilon$, on generation capacity factor $f$. Circles indicate simulation data and the curves give the theoretical result of Eq.~\eqref{S-opt}.  $L$ is  the daily load.  }
\label{theoryvssim}
\end{figure}

Equation~(\ref{S-opt}) illustrates the two key features of the tradeoff between generation and storage capacities: (a) The storage $S$ depends logarithmically on $\varepsilon$; thus a small increase in storage capacity substantially increases the combined system reliability. (b) A small increase in $f$ beyond 1  substantially decreases the required storage capacity (Fig.~\ref{theoryvssim}). 

\section{Cost Optimization}
\label{sec:cost}

We now determine the optimal configuration of the combined system by minimizing the cost function
\begin{align}
    {\cal C} = C_s\, \frac{S(f)}{L}+ C_g \,f\,.
    \label{totalcost}
\end{align}
Here again $C_g=\$75$ billion is the cost of a solar farm whose area $A_0$ is just sufficient to supply the daily electrical load $L$ of the St.\ Louis region during an average insolation day at the winter solstice, while  $C_s=\$22$ billion is the cost of a storage system that can supply one day of electrical load for the region. As mentioned previously, the cost of the generation system is assumed to be linear in its area, so the cost of a solar farm of area $fA_0$ will be $C_g f$.  Similarly, a storage system that supplies an energy $S$  will have a cost $C_s S/L$, under the assumption that the cost of storage is also linear in its capacity.

To find the optimal parameters $(f^*,S^*)$ in the minimum cost configuration, we set $d\mathcal{C}/df=0$ to give
\begin{align}
    \frac{dS(f)}{df}\Big|_{f^*,S^*} = -L\,\frac{C_g}{C_s}.
    \label{dsdf}
\end{align}
Thus the optimal system configuration depends only on the ratio of storage to generation cost, $C_s/C_g$, once $\varepsilon$ is specified. (The relation between the ratio $C_s/C_g$ and conventional measures of storage and generation costs is given in Appendix~\ref{sec:SM-costratio}.)~ The details of this minimization are given in Appendix~\ref{sec:SM-fstar}, from which the optimal solar farm size is determined from
\begin{align}
    f^* = 1+ \frac{2 \lambda_0}{\Gamma} \frac{C_s/(C_gr_0)-1}{\sqrt{2C_s/(C_gr_0)-1}}\,,
    \label{fexact}
\end{align}
where the dimensionless parameter $r_0$ is given by 
\begin{align}
r_0 = \frac{2\lambda_0^2 L}{\Gamma \ln{(\varepsilon_0/\varepsilon})}~.
\label{rzero}
\end{align}
The optimal storage value $S^*$ is then obtained by substituting $f^*$ in Eq.~\eqref{S-opt}.

Here, and in what follows, we use $\varepsilon=0.03$ (failure about once every 33 years) because it roughly corresponds to the accepted standard of a load loss of one day per ten years~\cite{standards}.  For this value of $\varepsilon$, $r_0 = 0.038$.  If the cost ratio $C_s/C_g$, which currently is roughly 0.3, were to become less than 0.038, then Eq.~\eqref{fexact} gives $f^*<1$ and indeed $f^*$ would not be defined if $C_s/C_g$ became less than 0.019. In this regime, our theory no longer applies, but is also unlikely to be reached by reductions in storage cost in the foreseeable future.

\begin{figure}[ht]
\centerline{\includegraphics[width=0.4\textwidth]{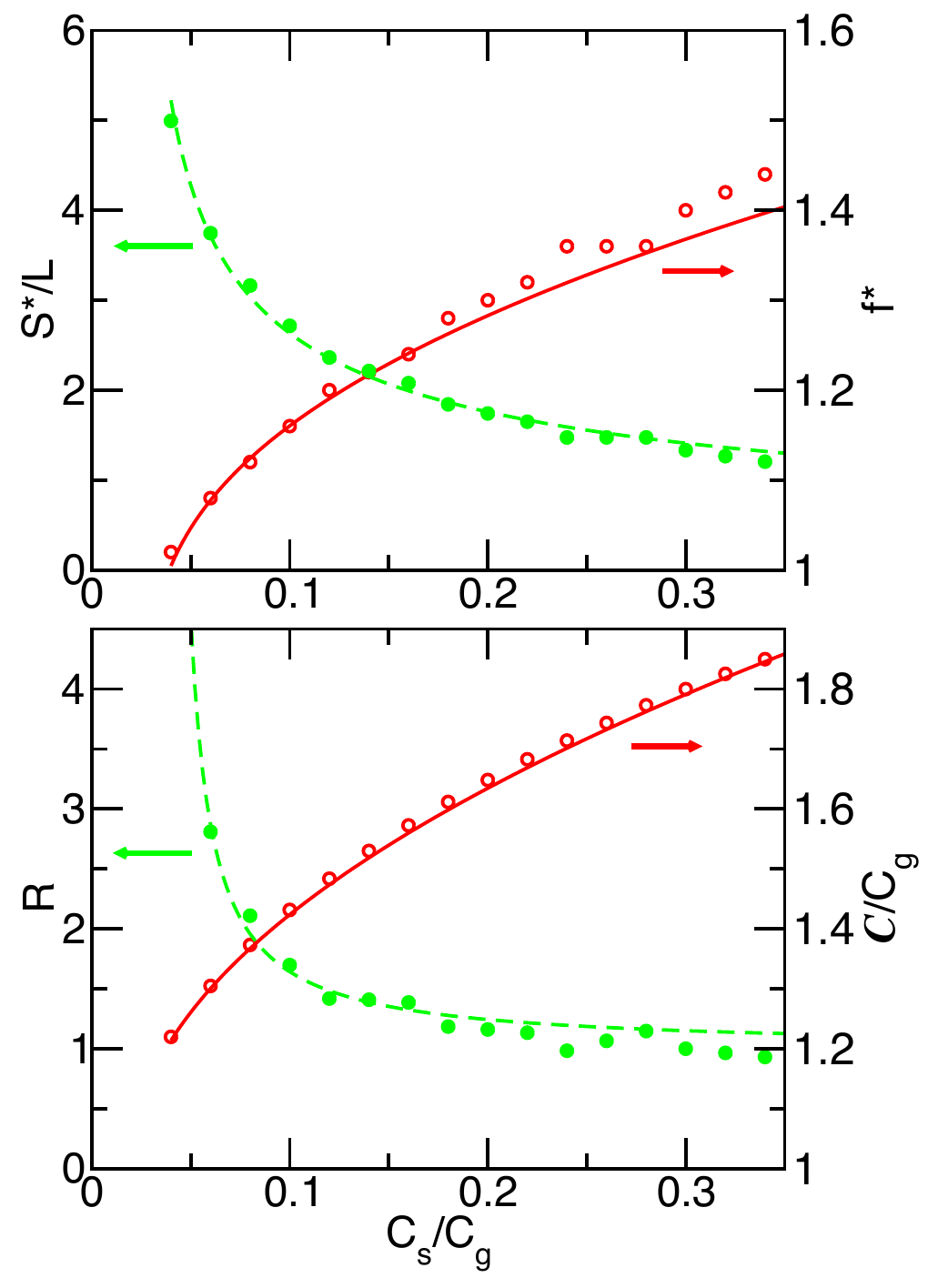}}
\caption{(top) Optimal storage value $S^*$ and generation capacity $f^*$, as functions of cost ratio $C_s/C_g$. (bottom) Ratio $R$ of storage cost to excess generation cost, and the system cost $\mathcal{C}$, as functions of $C_s/C_g$. In both panels, circles are simulation points, while solid lines are the theoretical predictions of Eqs.~(\ref{S-opt}) and (\ref{fexact}).}
\label{fig:costpicture}
\end{figure}

Our theoretical predictions for $(f^*,S^*)$ agree well with simulation results shown in Fig.~\ref{fig:costpicture}. Over the range of cost ratios shown, $S^*$ varies from 1 to 7 days, while $f^*$ varies from near 1 to 1.4. To gain insight into the dependences of $(f^*,S^*)$ on system parameters, it is helpful to focus on the limits of expensive and inexpensive storage:

\subsection{Expensive storage limit} We define this limit as $C_s/(C_g r_0)\gg 1$, where \eqref{fexact} reduces to 
\begin{subequations}
\begin{align}
    f^* \simeq 1+\frac{ \lambda_0}{\Gamma} \sqrt{\frac{2C_s}{C_g r_0}} = 1+\sqrt{\frac{\ln{(\varepsilon_0/\varepsilon)}}{\Gamma L} \frac{C_s}{C_g}}\,.
    \label{flarge}
\end{align}
When storage is expensive, the combined system favors generation over storage. Consequently, $f$ becomes large, so that we can drop the $\lambda_0$ term in Eq.~\eqref{S-opt} to give
\begin{align}
    S^*  \simeq \frac{\ln \left(\varepsilon_0/\varepsilon  \right)}{\Gamma(f^*-1)} 
= \sqrt{\frac{\ln{(\varepsilon_0/\varepsilon)}L}{\Gamma} \frac{C_g}{C_s}}\,.
   \label{slarge}
\end{align}
\end{subequations}
Thus $f^*$ and $S^*$ have inverse dependences on the cost ratio $C_g/C_s$. Combining Eqs.~(\ref{flarge}) and (\ref{slarge}), the ratio $R$ of the \emph{total} storage cost to the \emph{excess} generation capacity cost is particularly simple:
\begin{align}
R\equiv \frac{C_s S^*/L}{C_g (f^*-1)}=1\,.
\label{equalcosts}
\end{align}
As shown in Fig.~\ref{fig:costpicture}, this ratio is already close to 1 when $C_s/C_g$ exceeds 0.2.

From Eqs.~(\ref{totalcost}) and (\ref{equalcosts}), we may write the total system cost in the equivalent forms
\begin{align}
    \mathcal{C}&= C_g + 2 C_s\frac{S^*}{L} = C_g + 2 C_g(f^*\!-\!1) \nonumber\\[2mm]
&= C_g  + 2\sqrt{  \frac{\ln (\varepsilon_0/\varepsilon)C_gC_s }{\Gamma L } } \,.
\label{totalcostlarge}
\end{align}
The first term in each of these forms is the ``bare'' cost of the generation system that would be adequate in the absence of insolation fluctuations. The second term represents the additional system cost that is needed to mitigate the effect of fluctuations. For $\varepsilon=0.03$ and $C_s/C_g$ in the range $[0.1,0.3]$, this additional cost is roughly 50--80\% of $C_g$ or \$40--\$60 Billion.  Equation~(\ref{totalcostlarge}) also shows that the additional system cost due to fluctuations increases only as $\sqrt{\ln(1/\varepsilon)}$. Thus, for example, to reduce $\varepsilon$ from $0.03$ to $0.003$, the excess cost needs to be increased by less than 15\%. 

Equation~(\ref{totalcostlarge}) also provides an explicit way to decide whether it is more cost effective to invest in reducing the generation cost or the storage cost. The quantity $C_s (\partial {\mathcal C}/\partial C_s)$ gives the sensitivity of the system cost to a given fractional reduction in $C_s$, while $C_g (\partial {\mathcal C}/\partial C_g)$ plays a similar role for $C_g$. From \eqref{totalcostlarge}, we find
\begin{align}
    \frac{C_s  (\partial {\mathcal C}/\partial C_s)}{C_g (\partial {\mathcal C}/\partial C_g)} = \frac{\sqrt{\ln{(\varepsilon_0/\varepsilon)}C_sC_g/\Gamma L}}{C_g+ \sqrt{\ln{(\varepsilon_0/\varepsilon)}C_sC_g/\Gamma L}  }.
    \label{costsensitivitylarge}
\end{align}
This cost sensitivity ratio is about $0.3$ when $C_s/C_g=0.3$. Thus a 30\% reduction in storage cost has about the same impact as a 10\% reduction in generation cost.

\subsection{Inexpensive storage limit}
We define this limit by $C_s/(C_g r_0)\! -\!1\ll 1$. Expanding Eq.~\eqref{fexact} and the denominator of Eq.~\eqref{S-opt} to first order in this quantity, we obtain
\begin{subequations}
\begin{align}
\label{fstarsmall}
f^*  & \simeq 1+\frac{2 \lambda_0}{\Gamma} \left( \frac{C_s}{C_g r_0}\! -\! 1 \right) 
 = 1\!-\!\frac{2 \lambda_0}{\Gamma} \!+\! \frac{\ln{(\varepsilon_0/\varepsilon)}}{\lambda_0 L}\frac{C_s}{C_g} \\
\label{sstarsmall}
S^* &= \frac{2 \lambda_0 L}{\Gamma} \frac{C_g}{C_s}
\end{align}
\end{subequations}
As $C_s/C_g$ approaches $r_0$, $f^*\to 1$ while $S^*$ approaches a constant value, so the excess system cost becomes dominated by the storage cost, as shown in Fig.~\ref{fig:costpicture}.

Combining (\ref{fstarsmall}) and (\ref{sstarsmall}), the total system cost is now
\begin{align}
  \mathcal{C} &\simeq C_g  + \frac{\ln (\varepsilon_0/\varepsilon)C_s}{\lambda_0 L}\,.
    \label{totalcostsmallf}
\end{align}
For $C_s/C_g = 0.04$, which is the smallest cost ratio value that we simulated, the additional cost due to weather fluctuations, (${\cal C} -C_g$), is roughly $\frac{1}{4}C_g$. Here, the additional system cost due to fluctuations increases as the logarithm of the inverse failure probability. Thus to reduce $\varepsilon$ from $0.03$ to $0.003$ requires a significant increase in the additional cost of about 30\%. 

The relative influence of cost reductions in storage versus generation is similar to that in the expensive-storage limit. The cost sensitivity ratio now becomes
\begin{align}
        \frac{C_s  (\partial {\mathcal C}/\partial C_s)}{C_g (\partial {\mathcal C}/\partial C_g)} = \frac{\ln{(\varepsilon_0/\varepsilon)}}{\lambda_0 L} \frac{C_s}{C_g},
    \label{costsensitivitysmall}
\end{align}
which is about $0.2$ at $C_s/C_g = 0.04$. Thus a 50\% reduction in storage cost now has about the same impact as a 10\% reduction in generation cost. In both the limits of expensive and inexpensive storage, reducing the generation cost has more impact on the overall cost than reducing storage cost.

\section{Discussion}

We developed an analytic theory to determine the optimal mix of solar generation and storage that minimizes the overall system cost and achieves a given reliability.  This system is specified by $f^*$, the ratio of the solar farm area to the area of a farm that fully supplies the electrical load $L$ for the St.\ Louis region on the \emph{average} minimum insolation day, and $S^*$ the capacity of the storage system, measured in units of daily load. Our modeling extends the work of Refs.~\cite{BucciarelliSE84,BucciarelliSE86,GordonSE87} by including seasonal insolation variations, a more realistic distribution of daily energies, and day-to-day correlations in the insolation. Based on a quasi-steady-state approximation for the fill level of the storage system, we obtained the following key results:

\begin{itemize}
    
\item The dependence of the failure probability on storage capacity $S$ (Eq.~\eqref{epsfinal}). By ignoring seasonal variations, Refs.~\cite{BucciarelliSE84,BucciarelliSE86,GordonSE87} found that the failure probability $\varepsilon$ decays exponentially in $S$. We find that that including seasonal variations gives a weak correction to this result. The near-exponential dependence of $\varepsilon$ on $S$ suggests that the seasonal variation is slow enough that the steady-state form that we used for the storage distribution $P(s)$ is accurate. 

\item The dependence of the storage required to achieve a given reliability, on the insolation fluctuations and and the generation capacity. We find that the storage need is an increasing function of daily insolation fluctuations, since they reduce $\Gamma$ in Eq.~\eqref{S-opt}. Without excess generation capacity ($f\!=\!1$), storage of almost a week of load is required to achieve a failure probability $\varepsilon$ less than $0.03$ (Fig.~\ref{theoryvssim}). The required storage decreases rapidly when $f$ increases from $1$. The functional form that we obtain for the decrease (Eq.~\eqref{S-opt}) differs from both the logarithmic dependence found in Ref.~\cite{GordonSE87} and the exponential one found in Ref. \cite{LorenzoSEM92}.

\item The cost of the optimal generation/storage system (Eqs.~(\ref{totalcostlarge}) and (\ref{totalcostsmallf})). 

\item A given percent reduction in the generation cost reduces the system cost by three to five times more than the same percent reduction in the storage cost  (Eqs.~\eqref{costsensitivitylarge} and \eqref{costsensitivitysmall}.)

\end{itemize}

A fundamental ingredient in our cost calculations is the ratio of the cost $C_g$ for a solar farm that can supply the daily load of the St.\ Louis region on an \emph{average} insolation day at the winter solstice, to the cost $C_s$ of storing one day of energy load. With current technology, this cost ratio, $C_s/C_g$, is roughly 0.3. From Fig.~\ref{fig:costpicture}, the optimal configuration is then given by $(f^*,S^*)\approx (1.4,1.3)$, which implies an overall system cost of $1.4\,C_g+1.3C_s+0.6 C_s \approx\$147$ Billion (where the last term incorporates the diurnal storage need), consistent with Ref.~\cite{tong2020effects}. As the storage cost decreases, the optimal generation capacity also decreases until the limiting case of $(f^*,S^*)\approx (1.0,5)$, with overall system cost $\approx$ \$91 Billion, is reached after a 7-fold decrease in storage.  If storage costs are smaller still, the optimal value $f^*$ becomes less than one, a range where our theory is not valid. Below we outline an approach to treat the range $f\ltwid 1$. 

A system cost of roughly \$100 Billion seems staggering. However, we emphasize that the long-term cost of a solar/storage system is likely cheaper than natural gas power generation.  The construction cost for the requisite 5 GW of natural gas generation for the St.\ Louis region is roughly \$4--5 Billion~\cite{const-cost1,const-cost2}. Based on prices in the recent past, the fuel cost per year of operation is about \$2.5 Billion~\cite{gas-cost1,gas-cost2}. However, gas prices have increased by a factor of three recently~\cite{gas-cost3}. Thus, assuming a 20-year amortization, the cost of natural gas generation will lie between \$55 Billion, based on the average gas price in the previous decade, and \$155 Billion, using the current price. The renewable-energy systems modeled here will be cheaper if the gas price is at the upper end of the range. This finding is consistent with that of Ref.~\cite{tong2020effects}, while Refs.~\cite{JacobsonPNAS15} and \cite{jacobson2022low} found renewable systems to be even more cost-effective. Our estimates neglect maintenance costs, but we anticipate that maintenance of solar/storage will be cheaper than that of natural gas because the former has almost no moving parts. The primary impediment to implementing a solar/storage system is its huge upfront capital cost.

Within a 100\% renewable system, costs can be reduced by deploying a mix of solar and wind energy~\cite{CaldeiraEES18,BofingerRE10}. Ref.~\cite{tong2020effects} found that such a mix would reduce costs by about 50\% relative to solar-only generation.  One advantage of solar/wind generation is that the wind is typically stronger when it is overcast, so an energy deficit in one mode of generation would be offset by a surplus in the other. 

If one is willing to forgo 100\% renewable energy generation, a solar/storage system could be augmented by natural gas ``peaker" plants that  operate only during solar energy deficit periods near the winter solstice.  Because natural gas generation plants are relatively cheap to build (as mentioned above), they are well suited to being run for just a few days of the year. Thus consider a composite system that consists of a solar farm of area $fA_0$, with $f\ltwid 1$, which is supplemented by a 5 GW natural gas peaker plant. 

In the absence of insolation fluctuations, the annual energy deficit for such a solar farm is approximately (see Appendix~\ref{sec:no-fluct}) 
\begin{align}
\label{DeltaE}
    \mathcal{D} = \frac{\sqrt{2}}{\pi} 365 L (1-f)^{3/2}\approx 164L\,(1-f)^{3/2}\,.
\end{align}
Assuming that the fluctuation contribution to the energy deficit is constant, and using a daily fuel cost of \$7 Million (the price over the past decade), the cost of a combined solar/peaker generation plant, amortized over the assumed lifespan of 20 years, is 
\begin{align}
    C&= \tfrac{1}{20}\; f C_g + \$7 \;\text{M}\times [164L\,(1-f)^{3/2}+10L] \nonumber \\
    &\to \$[\alpha f+\beta \,(1-f)^{3/2}]\; \text{Billion}\,,
\end{align}
with $\alpha =3.75$ and $\beta=1.15$. In the last line, we ignore the contribution that is independent of $f$. The crucial ingredient is $\beta/\alpha\approx 0.307$, the relative cost of natural gas to solar. With increasing $\beta$ (increasing gas price), the cost-optimal value of $f$ will increase. However, regardless of how expensive gas becomes, the cost-minimizing system will always have $f < 1$---in other words, some use of peakers will be cost-effective. The optimal combination of generation, storage, and peakers is a question to be determined by future analyses. Our analytic results for the failure probability and required storage will aid such efforts.

\begin{acknowledgments}
SR thanks Dan Shrag for helpful advice and conversations, and acknowledges partial financial support from National Science Foundation grants DMR-1910736 and EF-2133863.
\end{acknowledgments}

\newpage
\widetext

%\bibliography{storage}
\appendix
\renewcommand{\thefigure}{A\arabic{figure}}
\setcounter{figure}{0}

\newpage
\section{Glossary}
In spite of our minimalist theoretical modeling starting point, a large number of parameters unavoidably arise. For the convenience of the reader, the most important of these parameters and their numerical values are given in Table~\ref{tab-g}.
\begin{table*}[ht]
\bigskip
\caption{List of the essential parameters used in this work. }
\begin{tabular}{ l l l }
Symbol & Definition &  Value (if fixed)  \\
  & & \\
$A_0$ & Area of solar farm that meets the daily electrical load at the solstice & $2.5\times10^8 \,\text{m}^2$ \\
$f$ & Generation capacity relative to a solar farm of area $A_0$  &   \\
$f^*$ & Cost-optimized value of $f$  &   \\
$S$& Capacity of the storage system  &   \\
$S^*$ & Cost-optimized value of $S$ & \\
$s(t)$ & Amount of energy in storage at time $t$& \\
$L$ & Daily energy load of the St.\ Louis region& $1.1\times 10^8$ kWh $\approx 4\times 10^{14}$ J\\
$E_j$ & Incident energy per m$^2$ on $j^{\rm th}$ day of the year & \\
$\sigma_j$ & Standard deviation of $E_j$ & \\
$\widetilde{\sigma}_j$ & Scaled standard deviation of $E_j$ & \\
$E$ & Value of $E_j$ at minimum insolation, smoothed over 45 days & 7.95 MJ/m$^2$ \\
$\sigma$ &   Value of $\sigma_j$ at minimum insolation, smoothed over 45 days & $2.79~\text{MJ/m}^2$ \\
$\widetilde{\sigma}$ &   Scaled value of $\sigma$ & 0.351 \\
$q$ & Day-to-day correlation parameter of solar insolation & 0.6157\\
$P_j(s)$ & Probability that the storage system contains energy $s$ on the $j^{\rm th}$ day of the year & \\
$P_{\rm min}(s)$ & Value of $P_j (s)$ on the minimum-insolation day &  \\
$\lambda_j$ & Decay constant in $P_j(s)\sim e^{\lambda_j s}$ & \\
$\lambda_{\rm cb}$ & Value of $\lambda_j$ for constant bias & \\
$\lambda_{\rm min}$ &  Value of $\lambda_j$ on the minimum-insolation day & \\
$\lambda_0$ & Value of $\lambda_{min}$ for $f=0$ & $1.055/L$ \\
$\Gamma$ & Coefficient in the linear interpolation $\lambda_{min}=\Gamma\times (f-1)$ &  $10.1/L$ \\
$C_g$ & Cost of generation system to supply the daily load at the insolation minimum  & \$75 Billion \\
$C_s$ & Cost of storage system that can store the daily load   & \$22 Billion \\
$\mathcal{C}$ & Total system cost & \\
$\varepsilon$ & Annual failure probability of the generation/storage system\\
$\varepsilon_j$ & Failure probability on the $j^{\rm th}$ day of the year\\
$\varepsilon_0$ & Eq.~(F4), prefactor for the failure probability & 9.72\\
$r_0$   & Eq.~(\ref{rzero}), characteristic cost ratio scale & 0.038 (for $\varepsilon= 0.03)$\\
$R$ & Eq.~(\ref{equalcosts}), ratio of storage cost to excess generation cost\\
\label{tab-g}

\end{tabular}
\end{table*}

\section{Calculation of Weather Statistics}
\label{sec:SM-ws}

The weather data was taken from the MERRA-2 dataset~\cite{molod2015development} that 
covers the period 1980 through 2019, using the ``SWGDN'' variable~\cite{merralink} (surface incoming shortwave flux, in W/m$^2$). This dataset supplies hourly solar resource data on a $0.5^{\circ}$ $\times$ $0.625^{\circ}$ mesh. We used a subset of $7\times 5$ mesh points surrounding St.\ Louis to define the energy incident on the St.\ Louis region. For each day, the hourly data was added over these $7\times 5$ mesh points to obtain a total daily energy.

\section{\protect Linear Interpolation Solution for $\lambda$ in Eq.~(3)} 
\label{sec:SM-linear}

\begin{figure}[ht]
\centerline{\includegraphics[width=0.375\textwidth]{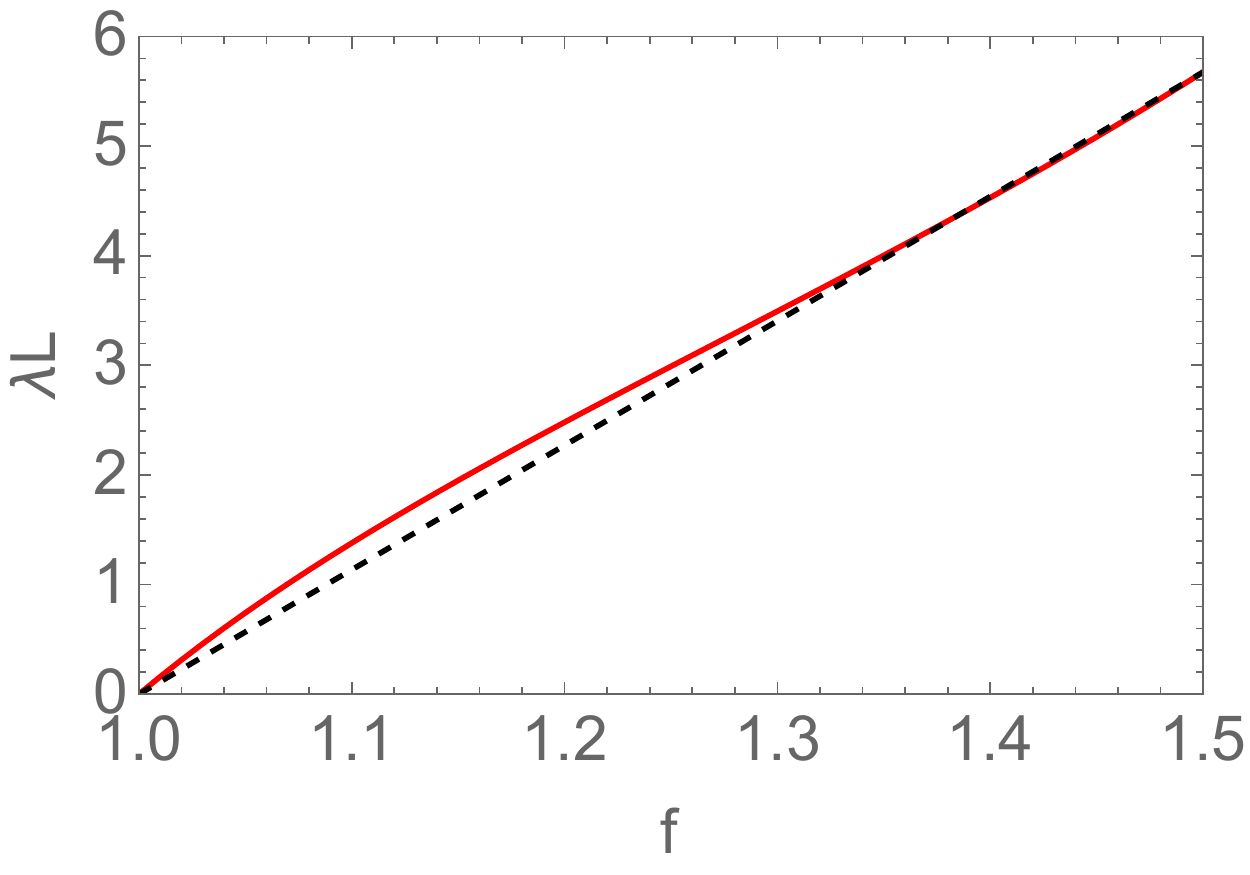}}
\caption{Exact numerical solution of Eq.~\eqref{sinh} (red) and the linear interpolation used in our theory (dashed).}
\label{fig:sinh}  
\end{figure}

To investigate the nature of the solution to Eq.~\eqref{sinh}, we define $z\equiv \sqrt{3} f\widetilde{\sigma} \lambda L$ and $\alpha\equiv(f-1)/\sqrt{3}f \widetilde{\sigma}$.  This transforms \eqref{sinh} to 
\begin{align}
\label{sh-sm}
\text{sinh}(z)/z =e^{\alpha z}\,.
\end{align} 
For solving this equation numerically, we use the values $L=1.1\times 10^8$ kWh and $\widetilde{\sigma} = 0.351L$. The numerical solution of this equation is shown in Fig.~\ref{fig:sinh}. For $f-1\to 0$ or equivalently, for $\alpha \to 0$, the solution for $z$ depends linearly on $f-1$.  For $f$ in the range 1.1--1.3 the dependence qualitatively deviates from strict linearity, but quantitatively remains close to the linear interpolation in the range $1<f<1.5$ shown in Fig.~\ref{fig:sinh}. This interpolation is obtained by matching the linear function with the exact numerical solution of \eqref{sh-sm}, $\lambda=5.675\ldots$, at $f=1.5$.

\section{Distribution of Storage at the Solstice}
\label{sec:SM-storage}
\begin{figure}[ht]
\centerline{\includegraphics[width=0.4\textwidth]{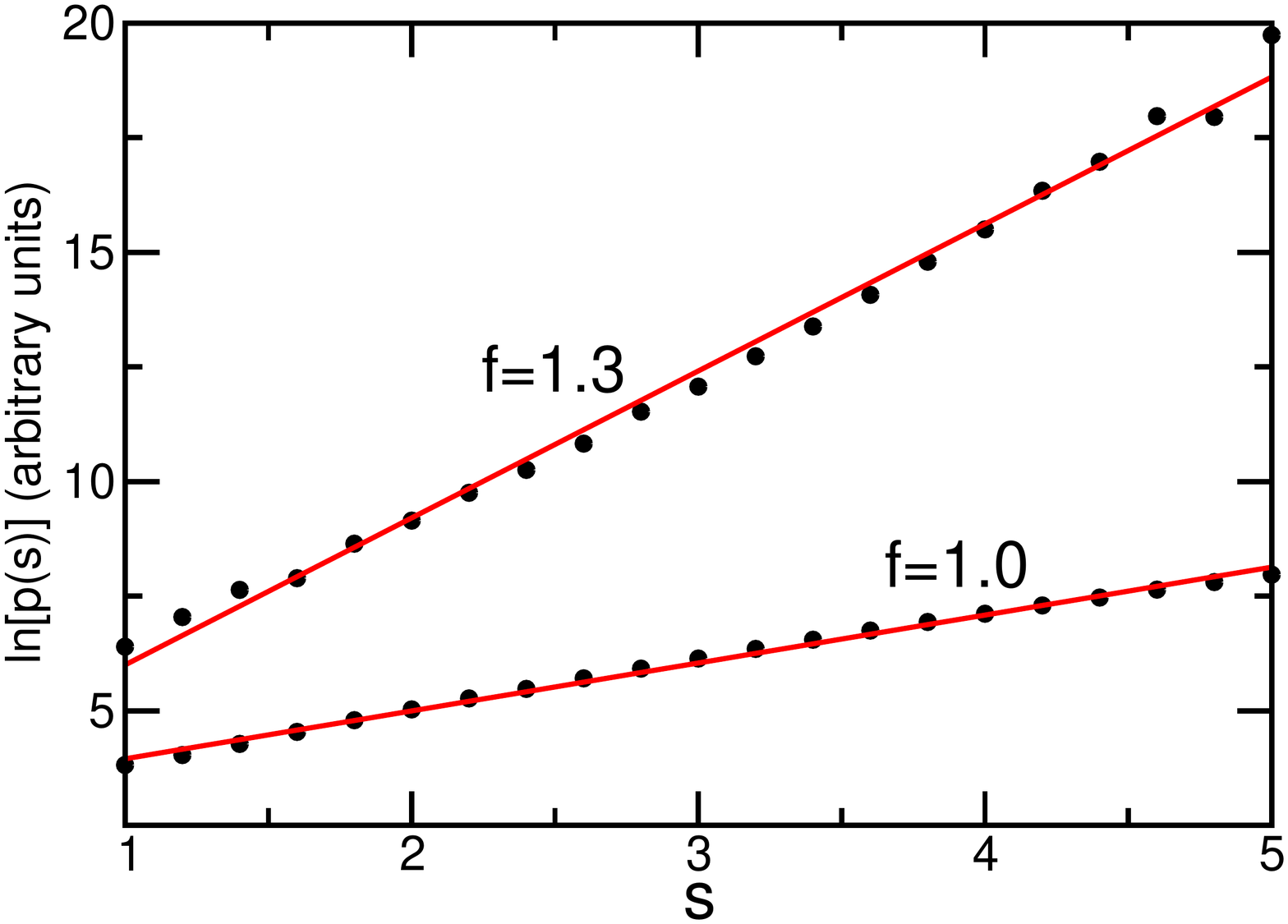}}
\caption{The distribution of stored energy on the minimum-insolation day, averaged over $10^7$ years, on a semi-logarithmic scale.}
\label{pofsplot}
\end{figure}

The theory outlined in Sec.~\ref{subsec:stored} ostensibly applies to the limit of strong bias, namely $f-1$, not small. In contrast, our simulations indicate that the storage distribution still has a nearly exponential form even when $f=1$, with a non-zero decay constant $\lambda_0$, whose numerical value is $1.140/L$ (Fig.~\ref{pofsplot}). This value of $\lambda_0$ is close to the value of $1.055/L$  that we obtained by the fitting procedure outlined in Sec.~\ref{subsec:stored}. This latter numerical value of $\lambda_0$ is what we used in our analytic theory.

\section{The Failure Probability}
\label{sec:SM-fp}
We compute the integral in Eq.~\eqref{failure} for the day-specific failure probability, $\varepsilon_j$, in which we use the approximate normalization constant $A_j\approx \lambda\, e^{-\lambda_j S}$ for $P_j(s)$.  We thus obtain
\begin{align}
\label{failure-SM}
   \varepsilon_j &= \frac{\mathcal{N}}{2\sqrt{3}f\widetilde{\sigma}_j}L \int_0^{\delta s} (e^{\lambda_j s}-1) (\delta s -s)\,ds \nonumber\\[2mm]
   & \approx \frac{e^{-\lambda_j S}}{2\sqrt{3} f\lambda\widetilde{\sigma}_j L\lambda} 
   \left(e^{\lambda_j \delta s} - 1-\lambda_j \delta s - \lambda_j^2\, \delta s^2/2
   \right)\nonumber\\[2mm]
   &\equiv A_j\,e^{-\lambda_j S}\,.
\end{align}   

Next we compute the annual failure rate $\varepsilon$ by adding the day-specific failure probabilities over a range of days around the winter solstice.  The first step in this calculation is to convert the discrete sum to an integral:
\begin{align}
    \varepsilon =\sum_j \varepsilon_j& =\sum_j A_j\,e^{-\lambda_j S}\nonumber
    \simeq\int_{-\infty}^{\infty}  A(t)\, e^{-\lambda(t,f)S}\,dt \,.
\end{align}
In the integral, we replace the discrete index $j$ by a dimensionless time variable $t$ that is expressed in units of days, in both the constant $A$ and in the decay rate $\lambda$.  Thus the integral is dimensionless, as it must be. To evaluate the integral, we exploit the fact that the amplitude $A_j$ is slowly varying with $j$, while the Gaussian factor (see Eq. \ref{lambdatf}) quickly diminishes for times away from the solstice.  Thus we make the approximation in which we evaluate $A_j$ on the minimum insolation day and define this value as $A_{\rm min}$.  We then take this amplitude outside the integral, and use Eq.~\eqref{lambdatf} to write the dependence of the decay rate on each day to give
\begin{align}
\varepsilon  &\simeq A_{\rm min}\int_{-\infty}^{\infty} e^{-\lambda_{\rm min}(f)S - f\,\Gamma\,S (t-t_{\rm min})^2/t_0^2}\;dt\nonumber
    =\frac{B}{\sqrt{\lambda_{\rm min}(f) S}}\;e^{-\lambda_{\rm min}(f)\,S},
    \label{epsfinal-SM}
\end{align}
where 
$B = A_{\min}\; t_0 \sqrt{\pi \lambda_{\rm min}(f)/f \Gamma }$.

\section{Solution to Eq.~(7)}
\label{sec:SM-S}
We want to invert \eqref{epsfinal}, $\varepsilon = B\, e^{-\lambda_{\rm min}S}/\sqrt{\lambda_{\rm min}S}$, and solve for $S$ as a function of $\varepsilon$.  By elementary steps, we rewrite this equation as 
\begin{align}
    2\left(\frac{B}{\varepsilon}\right)^2= w\, e^w\,,
\end{align}
where $w = 2\lambda_{\rm min} S$.  The solution for $w$ in this equation is the Lambert function. In the relevant limiting case in which $B/\varepsilon\gg 1$, $w$ has the asymptotic behavior
\begin{align}
    w= 2\lambda_{\rm min} S \simeq \ln \left[ 2 (B/\varepsilon)^2\right] -  \ln\left\{\ln \left[ 2 (B/\varepsilon)^2\right] \right\}\,.
\end{align}
Solving for $S$ and neglecting higher-order terms then gives
\begin{align}
\label{loglog}
    S\simeq \frac{\ln(B/\varepsilon)}{\lambda_{\rm min}} -\frac{\ln\ln(B/\varepsilon)}{2\lambda_{\rm min}}\,. 
\end{align}
Because the double logarithm in Eq.~\eqref{loglog} is very slowly varying in $\varepsilon$, we replace the factor $\varepsilon$ in this double logarithm by a constant that is equal to its value at $\varepsilon=0.003$, which is in the middle of the range of reliability, $0.0003\leq\varepsilon\leq 0.03$ that we investigated numerically.  Using this last simplification, we rewrite \eqref{loglog} as
\begin{align}
    S =\frac{\ln{(\varepsilon_0/\varepsilon)}}{\lambda_{\rm min}},
\end{align}
where $\varepsilon_0 = {B}/\sqrt{\ln{B/0.003}}$.
This result, in combination with Eq. \eqref{lambdamin}, gives Eq.~\eqref{S-opt} in the main text.

\section{\protect Calculation of $C_s/C_g$ from standard cost measures}
\label{sec:SM-costratio}

The cost of energy generation and storage is typically quoted as dollars per watt of generation capacity and dollars per kWh of capacity for storage. Currently, the generation cost $c_g$ is roughly \$1.50 per Watt and the storage cost $c_s$ is roughly \$200 per kWh. On the other hand, in this manuscript we have quoted generation and storage costs, each of a specified size, in units of dollars only.  Here, we relate the cost ratio $C_s/C_g$ used in this manuscript to the more conventional cost ratio $c_s/c_g$. This relation can be used to calculate the ratio $C_s/C_g$ in different geographical regions, and does not depend on assumptions about solar panel efficiency. 

Since the generation capacity of solar panels is given in watts, we have
\begin{align}
C_g = G\,c_g
\label{defcg}
\end{align}
where $G$ is the generation capacity in watts required to satisfy the daily load of the St.\ Louis region at the solstice and $c_g$ is the cost of the panels per watt. The generation capacity $G$ is defined with respect to a reference insolation intensity of $1000\,\text{W/m}^2$. Consequently, the power generated at an insolation intensity $I$ is $G \times (I/1000\,\text{W/m}^2$).  Thus if the generation capacity $G$ generates an energy $L$ in a day at the solstice, $L=G\times ({\bar I}/1000\,\text{W/m}^{2})\times 86,400~\text{sec}$, where $\bar I$ is the  day-averaged insolation level at the solstice and 1 day = 86,400 sec. Then we have
\begin{align}
G  = \frac{L}{{86,400~\text{sec}}} \frac{1000\,\text{W/m}^{2}}{{\bar I}}\;.
\label{capacity}
\end{align}

On the other hand $c_s$ is given in dollars per kWh, so that $C_s = L(\text{in}~\text{kWh})\times c_s$. Combining this with Eqs.~(\ref{defcg}) and (\ref{capacity})  gives
\begin{align}
\frac{C_s}{C_g}  &= \frac{1W}{1kWh} (86,400 \text{sec}) \frac{\bar I}{1000\,\text{W/m}^{2}} \frac{c_s}{c_g} 
= 0.024 \frac{\bar I}{1000\,\text{W/m}^{2}} \frac{c_s}{c_g},
\label{csbycgnew1}
\end{align}
where we use $1~\text{kWh} = 3.6\times 10^6\, \text{J}$. For the St.\ Louis region, we use the data given in Sec.~\ref{subsec:data}, namely, ${\bar I} = (7.95\, \text{MJ/m}^2)/(86,400~\text{sec}) = 91\,\text{W/m}^2$, so that 
\begin{align}
\frac{C_s}{C_g}  = {0.0022} \frac{c_s}{c_g}\,.
\label{csbycgnew2}
\end{align}
For $c_s=$\$200/kWh and $c_g$=\$1.50/W, this gives $C_s/C_g\sim 0.3$.

\section{Derivation of Eq.~\eqref{fexact}}
\label{sec:SM-fstar}

We write Eq.~\eqref{S-opt} in the form
\begin{align}
    S= \frac{\left[ \ln{(\varepsilon_0/\varepsilon)}/\lambda_0 \right]} {x+\sqrt{1+x^2}} = \left[ \ln{(\varepsilon_0/\varepsilon)}/\lambda_0 \right](\sqrt{1+x^2} -x)\,,
\end{align}
where $x=\Gamma (f-1)/2\lambda_0$. Using the chain rule, Eq.~\eqref{dsdf}, and the definition $r_0 = (2\lambda_0^2 L)/\big[\Gamma \ln{(\varepsilon_0/\varepsilon})\big]$, we find
\begin{align}
   \frac{x}{\sqrt{1+x^2}}-1 =- \frac{2 \lambda_0 L}{\Gamma} \frac{\lambda_0}{\ln{(\varepsilon_0/\varepsilon)}} \frac{C_g}{C_s}= - r_0C_g/C_s\,.
\end{align}
Then 
\begin{align}
 \frac{x}{\sqrt{1+x^2}} = 1-r_0C_g/C_s   \,.
\end{align}
Squaring both sides and solving for $x$ gives 
\begin{align}
x = \frac{1-r_0C_g/C_s}{\sqrt{2r_0C_g/C_s-r_0^2 C_g^2/C_s^2}}
= \frac{C_s/C_g r_0-1}{\sqrt{2C_s/C_g r_0 -1}} \,.
\end{align}
Using the definition of $x$ then gives Eq.~\eqref{fexact}.

\section{Solar Energy Deficit Near the Winter Solstice}
\label{sec:no-fluct}

For a solar farm of area $fA_0$ with $f\ltwid 1$, with a deterministic sinusoidal solar insolation time dependence, we estimate the energy deficit over the course of a  year. Under these assumptions the daily energy produced by a solar farm of area $A$ is
\begin{align}
  \label{Et}
  E(t) = (\overline{E}+\Delta E\,\cos\omega t)f A_0\,e\,,
\end{align}
where $e$ is the efficiency of the solar panels. Reading off from Fig.~\ref{fig:av-E}, we infer the values $\overline{E}\approx 16$MJ/m$^2$ and $\Delta E\approx 8$MJ/m$^2$. For the purpose of a simple estimate, we take these numbers to be exact. 

\begin{figure}[ht]
  \centerline{\includegraphics[width=0.35\textwidth]{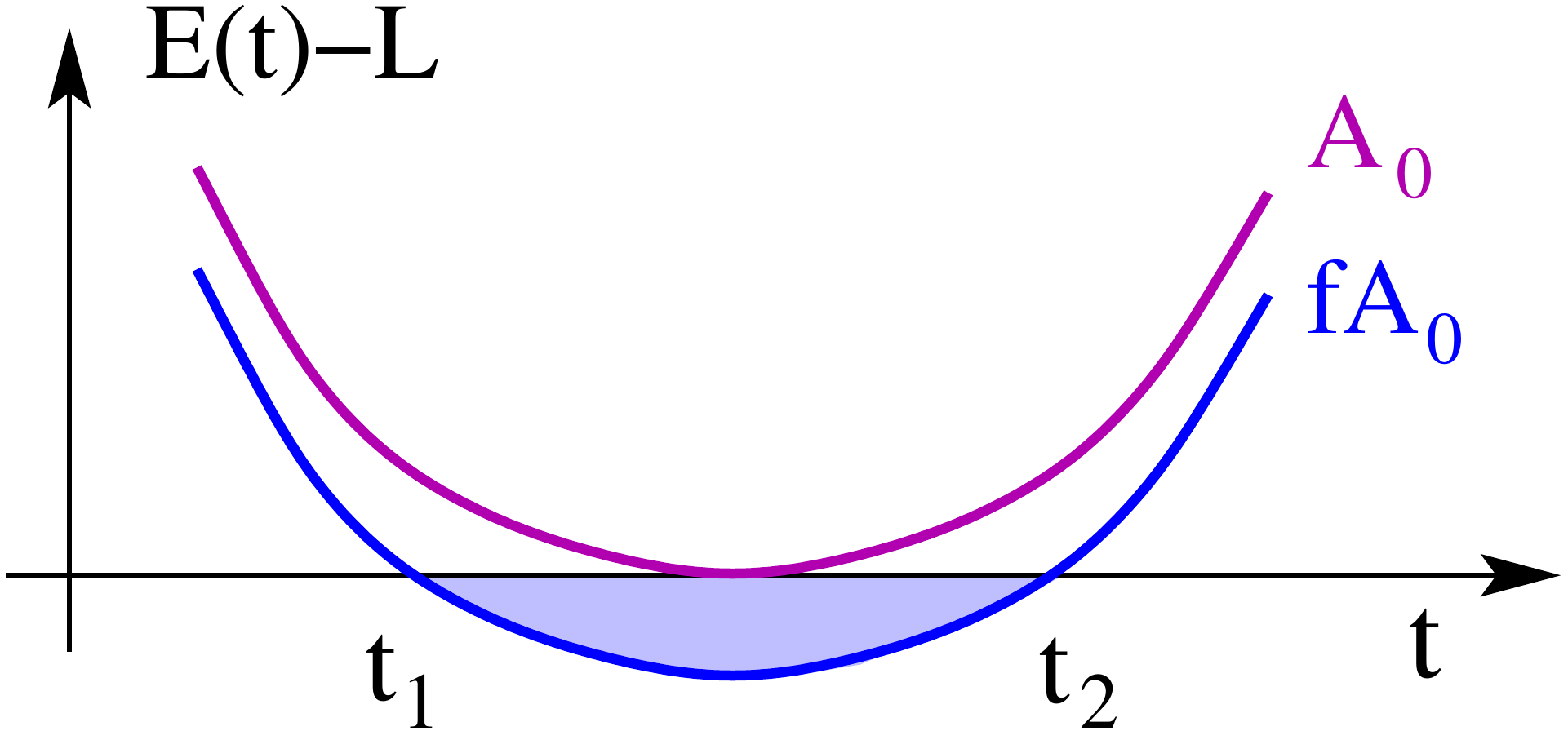}}
  \caption{Schematic time dependence of the daily solar energy minus the load for  sinusoidal insolation on a solar farm of area $A_0$ (blue) and of area $fA_0$ (red), with $f\ltwid 1$. The annual energy deficit $\mathcal{D}$. is shown shaded}
\label{fig:de-solstice}  
\end{figure}

The energy produced each day by a solar farm of area $fA_0$ on the $t^{\text th}$ day of the year is
\begin{align}
  E(t) = (\overline{E}+\Delta E\,\cos\omega t)fA_0\,e\,,  
\end{align}
and a shortage arises when $E(t)$ first becomes less than $L$.  This event happens at a time $t_1$ when
\begin{align}
  \label{Ec}
   (\overline{E}+\Delta E\,\cos\omega t)_1fA_0\, e=L\,,
\end{align}
which occurs before the winter solstice (Fig.~\ref{fig:de-solstice}). Subsequently, the energy deficit increases day by day until a time $t_2$ when $E(t)$ again exceeds $L$. 

Notice that for $f=1$ the solar energy produced at the solstice matches the electrical load; that is, $(\overline{E}-\Delta E\,)A_0\epsilon = L$. Using this in \eqref{Ec} gives
\begin{align*}
  (\overline{E}+\Delta E\,\cos\omega t_1)fA_0\epsilon =
  (\overline{E}-\Delta E\,)A_0\epsilon\,.
\end{align*}
Solving for the time when the shortfall first occurs gives
  \begin{align}
    \label{cos}
  \cos \omega t_1 = \frac{\overline{E}}{\Delta E}\left(\frac{1}{f}-1\right)-\frac{1}{f}\,.
\end{align}
If we again use $\overline{E}= 16$MJ/m$^2$ and $\Delta E=8$MJ/m$^2$, then \eqref{cos} gives $\cos\omega t_1= \frac{1}{f}-2$. For $f\ltwid 1$, we set $f=1-\delta$, with $\delta\ll 1$, and obtain $\cos\omega t_1 \approx -1+\delta$.

We now set $t_1=\frac{1}{2}T-u$, where $T=365$ is the number of days in a year, and we also shift the time origin so that $\frac{1}{2}T$ corresponds to the winter solstice.  Thus $u$ is the number of days before the solstice when a solar energy deficit exists, which will be small when $\delta\ll 1$.  With these assumptions,
\begin{align*}
  \cos\omega t_1 =\cos\left[ \frac{2\pi}{T}\left(\frac{T}{2}-u\right)\right]
                = \cos\pi \cos z \approx -1+ z^2/2\,,
\end{align*}
where $z\equiv 2\pi u/T$.  Since $\cos\omega t_1$ also equals $-1+\delta$, we have $u= T\sqrt{2\delta}/(2\pi)$, for the number of days before the solstice that a solar farm of area $fA_0$ cannot supply the daily energy needs. Between $t_1\equiv \frac{1}{2}T-u$ and $t_2\equiv \frac{1}{2}T+u$, or in the time range $\Delta t \equiv t_2-t_1\approx \frac{T}{\pi} \sqrt{2\delta} =
\frac{T}{\pi}\sqrt{2(1-f)}$, the daily solar energy is less than the energy
load for each day.  The annual energy deficit in this time range is (see
Fig.~\ref{fig:de-solstice})
\begin{align}
  \mathcal{D}= \int_{t_1}^{t_2} dt \left|E(t)-L\right|\,.
\end{align}
Substituting $E(t)$ from \eqref{Et}, eliminating $A_0$ by the relation
$(\overline{E}-\Delta E\,)A_0\epsilon = L$, and again using $\overline{E}= 16$MJ/m$^2$ and $\Delta E=8$MJ/m$^2$, we obtain
\begin{align}
  \label{S*}
  \mathcal{D}  = L \int_{t_1}^{t_2} dt \left|\frac{ (\overline{E}+\Delta E\,\cos\omega t)f}
     {\overline{E}-\Delta E} -1\right|\nonumber
    &\to  L\int_{t_1}^{t_2} dt\left|(2+\cos\omega t)f-1\right|\nonumber\\
    &\approx L\Delta t(1-f)= \frac{\sqrt{2}}{\pi}LT(1-f)^{3/2}\,.
\end{align}
In the integrand, we evaluated the trigonometric function to first order for $\omega t$ close to $\pi$, since both $t_1$ and $t_2$ are close to $\frac{1}{2}T$ for $f\ltwid 1$, to give the result \eqref{DeltaE}.

\end{document}